\documentclass[11pt,letterpaper]{article}

\usepackage{typearea}
\paperwidth 8.5in \paperheight 11in \typearea{16}

\usepackage{theorem,latexsym,graphicx}
\usepackage{amsmath,amssymb,enumerate}
\usepackage{xspace}
\usepackage{bm}
\usepackage{ifpdf}
\usepackage{color}
\usepackage[compact]{titlesec}
\usepackage{algorithm, algorithmic}
\usepackage{paralist}

\newcommand{\ignore}[1]{}

\definecolor{Darkblue}{rgb}{0,0,0.4}
\definecolor{Brown}{cmyk}{0,0.81,1.,0.60}
\definecolor{Purple}{cmyk}{0.45,0.86,0,0}

\newcommand{\mydriver}{hypertex} \ifpdf
 \renewcommand{\mydriver}{pdftex}
\fi
\usepackage[breaklinks,\mydriver]{hyperref}
\hypersetup{colorlinks=true,
            citebordercolor={.6 .6 .6},linkbordercolor={.6 .6 .6},%
citecolor=Darkblue,urlcolor=black,linkcolor=Darkblue,pagecolor=black}

\newcommand{\lref}[2][]{\hyperref[#2]{#1~\ref*{#2}}}


\newtheorem{theorem}{Theorem}[section]
\newtheorem{definition}[theorem]{Definition}
\newtheorem{proposition}[theorem]{Proposition}

\newtheorem{property}[theorem]{Property}
\newtheorem{lemma}[theorem]{Lemma}

\newtheorem{claim}[theorem]{Claim}

\newtheorem{corollary}[theorem]{Corollary}

\allowdisplaybreaks

\newenvironment{Myquote}{\par\begingroup
\addtolength{\leftskip}{1em} \rightskip\leftskip }{\par
\endgroup
}

\newenvironment{proof}{

\noindent{\bf Proof:}} {\hfill$\blacksquare$

}

\newcommand{\junk}[1]{}

\newcommand{\R}[0]{{\ensuremath{\mathbb{R}}}}

\newcommand{\Z}[0]{{\ensuremath{\mathbb{Z}}}}

\def\p {\ensuremath{\mathcal{P}}\xspace}

\def\a{\ensuremath{\mathcal{A}}\xspace}
\def\rs{\mathcal{R}}

\def\f{\ensuremath {\mathcal{F}}\xspace}

\def\ms{\ensuremath{\mathcal{M}}\xspace}

\def\opt{{\sf Opt}\xspace}
\def\aug{{\sf Aug}}
\def\cov{\ensuremath{\Pi}\xspace}
\def\rcov{{\sf Robust(\cov)}\xspace}

\def\mm{{\sf MaxMin}\xspace}
\def\mmp{{\sf MaxMin(\cov)}\xspace}

\def\fst{\ensuremath {\Phi_T}\xspace}
\def\snd{\ensuremath {{\sf Augment}_T}\xspace}

\newcommand{\sse}{\subseteq}
\newcommand{\I}{{\Omega}}

\newcommand{\eeminusone}{\frac{\mathrm{e}}{\mathrm{e}-1}}
\newcommand{\ts}{\textstyle}

\newcommand{\offline}{\alpha_{\sf off}}
\newcommand{\online}{\alpha_{\sf on}}
\newcommand{\optaug}{{\sf OptAug}}

\newcounter{note}[section]

\newcommand{\initOneLiners}{%
    \setlength{\itemsep}{0pt}
    \setlength{\parsep }{0pt}
    \setlength{\topsep }{0pt}
}
\newenvironment{OneLiners}[1][\ensuremath{\bullet}]
    {\begin{list}
        {#1}
        {\initOneLiners}}
    {\end{list}}

\newcommand{\Tstar}{\ensuremath{T^*}\xspace}
\newcommand{\Phistar}{\ensuremath{\Phi^*}\xspace}

\usepackage{times}

\begin{document}

\title{Robust and MaxMin Optimization under \\Matroid and Knapsack
  Uncertainty Sets\thanks{An extended abstract containing the results of
    this paper and of~\cite{GNR-k-rob} appeared jointly in
    \emph{Proceedings of the 37th International Colloquium on Automata,
      Languages and Programming (ICALP), 2010}.}}

\author{
Anupam Gupta\thanks{Computer Science Department, Carnegie Mellon
    University, Pittsburgh, PA 15213, USA. Supported in part by
    NSF awards CCF-0448095 and CCF-0729022, and an Alfred P.~Sloan
    Fellowship. Email: anupamg@cs.cmu.edu}
\and Viswanath Nagarajan\thanks{IBM T.J. Watson Research Center, Yorktown Heights, NY 10598, USA. Email:
viswanath@us.ibm.com} \and R. Ravi\thanks{Tepper School of Business, Carnegie Mellon University,
  Pittsburgh, PA 15213, USA. Supported in part by NSF grant
  CCF-0728841. Email: ravi@cmu.edu}
}
\date{}
\maketitle

\begin{abstract}
  Consider the following problem: given a set system $(U,\I)$ and an edge-weighted
  graph $G = (U, E)$ on the same universe $U$, find the set $A \in \I$ such
  that the Steiner tree cost with terminals $A$ is as large as
  possible---``which set in $\I$ is the most difficult to connect up?''
  This is an example of a \emph{max-min problem}: find the set $A \in
  \I$ such that the value of some minimization (covering) problem is as
  large as possible.

  In this paper, we show that for certain covering problems which admit
  good deterministic online algorithms, we can give good algorithms for
  max-min optimization when the set system $\I$ is given by a $p$-system
  or knapsack constraints or both. This result is similar to results for constrained
  maximization of submodular functions. Although many natural covering problems
  are not even approximately submodular, we show that one can use
  properties of the online algorithm as a surrogate for submodularity.

  Moreover, we give stronger connections between max-min optimization
  and two-stage robust optimization, and hence give improved algorithms
  for robust versions of various covering problems, for cases where the
  uncertainty sets are given by $p$-systems and $q$ knapsacks.
\end{abstract}

\section{Introduction}

\newcommand{\maxf}{\textsf{Max-$f$}\xspace}

Recent years have seen a considerable body of work on the problem of constrained submodular maximization: you are given
a universe $U$ of elements, a collection $\I \sse 2^U$ of ``independent'' sets and a submodular function $f: 2^U \to
\R_{\geq 0}$, and the goal is to solve the optimization problem of maximizing $f$ over the ``independent'' sets:
\begin{equation}
  \max_{S \in \I} f(S). \tag{\maxf}
\end{equation}
It is a classical result that when $f$ is a linear function and $(U, \I)$ is a matroid, the greedy algorithm solves
this exactly. Furthermore, results from the mid-1970s tell us that even when $f$ is monotone submodular and $(U, \I)$
is a partition matroid, the problem becomes NP-hard, but the greedy algorithm is a $\eeminusone$-approximation---in
fact, greedy is a $2$-approximation for monotone submodular maximization subject to \emph{any} matroid constraint.
Recent results have shed more light on this problem: it is now known that when $f$ is a monotone submodular function
and $(U, \I)$ is a matroid, there exists a $\eeminusone$-approximation algorithm. We can remove the constraint of
monotonicity, and also generalize the constraint $\I$ substantially: the most general results say that if $f$ is a
non-negative submodular function, and if $\I$ is a \emph{$p$-system},\footnote{A $p$-system is
  similar to, but more general than, the intersection of $p$ matroids;
  it is formally defined in \lref[Section]{subsec:framework-mat}} then one
can approximate \maxf to within a factor of $O(p)$; moreover, if $\I$ is the intersection of $O(1)$ knapsack
constraints then one can approximate \maxf to within a constant factor.

Given this situation, it is natural to ask: \emph{For which broad
  classes of functions can we approximately solve the \textsf{Max-$f$}
  problem efficiently?} (Say, subject to constraints $\I$ that form a
$p$-system, or given by a small number of knapsack constraints, or
both.) Clearly this class of functions includes submodular functions.
Does this class contain other interesting subclasses of functions which
are far from being submodular?

In this paper we consider the case of ``max-min optimization'': here $f$
is a monotone subadditive function defined by a minimization covering
problem, a natural subset of all subadditive functions. We show
conditions under which we can do constrained maximization over such
functions $f$.  For example, given a set system $(U, \f)$, define the
``set cover'' function $f_{SC}: 2^U \to \Z_{\geq 0}$, where $f(S)$ is
the minimum number of sets from $\f$ that cover the elements in
$S$. This function $f_{SC}$ is not submodular, and in fact, we can show
that there is no submodular function $g$ such that $g(S) \leq f_{SC}(S)
\leq \alpha\; g(S)$ for sub-polynomial $\alpha$. (See
\lref[Section]{sec:lbd}.) Moreover, note that in general we cannot even
evaluate $f_{SC}(S)$ to better than an $O(\log n)$-factor in polynomial
time.  However, our results imply $\max_{S \in \I} f_{SC}(S)$ can indeed
be approximated well.  In fact, the result that one could approximately
maximize $f_{SC}$ subject to a cardinality constraint was given by Feige
et al.~\cite{FJMM07}; our results should be seen as building on their
ideas. (See also the companion paper~\cite{GNR-k-rob}.)

At a high level, our results imply that if a monotone function $f$ is defined by a (minimization) covering problem, if
$f$ is subadditive, and if the underlying (minimization) covering problem admits good deterministic online algorithms,
then there exist good approximation algorithms for \maxf subject to $p$-systems and $q$ knapsacks.  (All these terms
will be made formal shortly.) The resulting approximation guarantee for the max-min problem depends on the competitive
ratio of the online algorithm, and $p$ and $q$. Moreover, the approximation ratio improves if there is a better
algorithm for the offline minimization problem, or if there is a better online algorithm for a fractional version of
the online minimization problem.

\paragraph{Robust Optimization.}
Our techniques and results imply approximation algorithms for covering
problems in the framework of robust optimization as well. In the robust
optimization framework, there are two stages of decision making. E.g.,
in a generic robust optimization problem, one is not only given a set
system $(U, \I)$, but also an inflation parameter $\lambda \geq 1$. Then
one wants to perform some actions in the first stage, and then given a
set $A \in \I$ in the second stage, perform another set of actions
(which can now depend on $A$) to minimize
\[ \text{(cost of first-stage actions)} + \max_{A \in \I} \lambda \cdot
\text{ (cost of second-stage actions)} \] subject to the constraint that
the two sets of actions ``cover'' the demand set $A \in \I$. As an
example, in robust set cover, one is given another set system $(U, \f)$:
the allowed actions in the first and second stage are to pick some
sub-collections $\f_1$ and $\f_2$ respectively from $\f$, and the notion
of ``coverage'' is that the union of the sets in $\f_1 \cup \f_2$ must
contain $A$. (If $\lambda > 1$, actions are costlier in the second
stage, and hence there is a natural tension between waiting for the
identity of $A$, and over-anticipating in the first stage without any
information about $A$.)

Note that robust and max-min problems are related, at least in one direction: if $\lambda = 1$, there is no incentive
to perform any actions in the first stage, in which case the robust problem degenerates into a max-min optimization
problem. In this paper, we show a reduction in the other direction as well---if one can solve the max-min problem well
(and if the covering problem admits a good deterministic online algorithm), then we get an algorithm for the robust
optimization version of the covering problem as well. The paper of Feige et al.~\cite{FJMM07} gave the first reduction
from the robust set-cover problem to the max-min set cover problem, for the special case when $\I = \binom{U}{k}$; this
result was based on a suitable LP-relaxation. Our reduction extends this in two ways: (a) the constraint sets $\I$ can
now be $p$-systems and $q$ knapsacks, and (b) much more importantly, the reduction now applies not only to set cover,
but to many sub-additive monotone covering problems (those with deterministic online algorithms, as mentioned above).
Indeed, it is not clear how to extend the ellipsoid-based reduction of~\cite{FJMM07} even for the Steiner tree problem;
this was first noted by Khandekar et al.~\cite{KKMS08}.

\paragraph{Our Results and Techniques.} Our algorithm for the max-min
problem is based on the observation that the cost of a deterministic
online algorithm for the underlying minimization covering problem
defining $f$ can be used as a surrogate for submodularity in certain
cases; specifically, we show that the greedy algorithm that repeatedly
picks an element maintaining membership in $\I$ and maximizing the cost
of the online algorithm gives us a good approximation to the max-min
objective function, as long as $\I$ is a $p$-system.

We also show how to reduce the problem of maximizing such a function
over the intersection of $q$ knapsacks to $n^{O(1/\epsilon^2)}$ runs of
approximately maximizing the function over a single partition matroid at
a loss of a factor of $q(1+\epsilon)$, or instead to
$n^{O(q/\epsilon^2)}$ runs of approximately maximizing over a different
partiton matroid at a loss of a factor of $(1+\epsilon)$---this
reduction is fairly general and is likely to be of interest in other
contexts as well. These results appear in \lref[Section]{sec:max-min}.

We then turn to robust optimization. In \lref[Section]{sec:gen-sets}, we
show that given a deterministic online algorithm for the covering
function $f$, and an approximate max-min optimization algorithm for $f$
over a family $\I$, we get an algorithm for two-stage robust version of
the underlying covering problem with uncertainty set $\I$---the
approximation guarantee depends on both the competitive ratio of the
online algorithm, as well as the approximation guarantee of the max-min
problem.

Note that we can combine this latter reduction (using max-min algorithms
to get robust algorithms) with our first reduction above (using online
algorithms to get max-min algorithms); in \lref[Section]{sec:combine}, we
give a more careful analysis that gives a better approximation than that
obtained by just naively combining the two theorems together.

Finally, in \lref[Section]{sec:lbd}, we show that some common covering
problems (vertex cover and set cover) give rise to functions $f$ that
cannot be well-approximated (in a mutliplicative sense) by any
submodular function, but still admit good maximization algorithms by our
results in \lref[Section]{sec:max-min}.

\subsection{Related Work}
\label{sec:related-work}

Constrained submodular maximization problems have been very widely studied~\cite{NWF78I,NWF78II,S04,CCPV07,V08,KST09}.
However, as we mention above, the set cover and vertex cover functions are far from submodular. Interestingly, in a
recent paper on testing submodularity~\cite{SV10}, Seshadhri and Vondrak conjecture that the success of greedy
maximization algorithms may depend on a more general property than submodularity; this work provides further
corroboration for this, since we show that in our context online algorithms can serve as surrogates for submodularity.

Feige et al.~\cite{FJMM07} first considered the $k$-max-min set cover subject to $\I = \binom{U}{k}$ (the
``cardinality-constrained'' case)---they gave an $O(\log m \log n)$-approximation algorithm for the problem with $m$ sets and $n$ elements. They also
showed an $\Omega(\frac{\log m}{\log\log m})$ hardness of approximation for $k$-max-min (and $k$-robust) set cover. The
results in this paper build upon ideas in~\cite{FJMM07}, by handling more general covering problems and sets $\I$. To
the best of our knowledge, none of the $k$-max-min problems other than min-cut have been studied earlier; note that the
min-cut function is submodular, and hence the associated max-min problem
can be solved using submodular maximization.

The study of approximation algorithms for robust optimization was initiated by Dhamdhere et al.~\cite{DGRS05,GGR06}:
they study the case when the scenarios were explicitly listed, and gave constant-factor approximations for several
combinatorial optimization problems. Again, the model with implicitly specified (and exponentially many) scenarios $\I$
was considered in Feige et al.~\cite{FJMM07}, where they gave an $O(\log m \log n)$-approximation for robust set cover
in the cardinality-constrained case $\I = \binom{U}{k}$. Khandekar et al.~\cite{KKMS08} noted that the techniques
of~\cite{FJMM07} did not seem to imply good results for Steiner tree, and developed new constant-factor approximations
for $k$-robust versions of Steiner tree, Steiner forest on trees and facility location, again for the
cardinality-constrained case. We investigate many of these problems in the cardinality-constrained case of both the
max-min and robust models in the companion paper~\cite{GNR-k-rob}, and obtain approximation ratios better than the
online competitive factors. On the other hand, the goal in this paper is to give a framework for robust and max-min
optimization under general uncertainty sets.

\ignore{Considering the \emph{average} instead of the worst-case performance gives rise to the well-studied model of
stochastic optimization~\cite{RS04, IKMM04}.  Some common generalizations of the robust and stochastic models have been
considered (see, e.g., Swamy~\cite{Swamy08} and Agrawal et al.~\cite{ADSY09}).}




\section{Preliminaries}
\label{sec:prelim}

\subsection{Deterministic covering problems}

A covering problem \cov has a ground-set $E$ of elements with costs $c:E\rightarrow \mathbb{R}_+$, and $n$ covering
requirements (often called demands or clients), where the solutions to the $i$-th requirement is specified---possibly
implicitly---by a family $\mathcal{R}_i \sse 2^E$ which is upwards closed (since this is a covering problem).
Requirement $i$ is \emph{satisfied} by solution $\f \sse E$ iff $\f\in \mathcal{R}_i$.  The covering problem $\cov =
\langle E,c, \{\mathcal{R}_i\}_{i=1}^n \rangle$ involves computing a solution $\f\sse E$ satisfying all $n$
requirements and having minimum cost $\sum_{e\in \f} c_e$.  E.g., in set cover, ``requirements'' are items to be
covered, and ``elements'' are sets to cover them with. In Steiner tree, requirements are terminals to connect to the
root and elements are the edges; in multicut, requirements are terminal pairs to be separated, and elements are edges
to be cut.

The min-cost covering function associated with \cov is:
$$f_\cov(S) := \min\left\{ \sum_{e\in \f} c_e \,:\,\, \f\in \mathcal{R}_i \mbox{ for all }i\in S\right\}.$$

\subsection{Max-min problems} Given a covering problem $\cov$ and a collection
$\Omega\sse 2^{[n]}$ of ``independent sets'', the {\em max-min} problem \mmp involves finding a set $\omega \in \Omega$
for which the cost of the min-cost solution to $\omega$ is maximized,
$$\max_{\omega\in\Omega}\,\,f_\cov(\omega).$$

\subsection{Robust covering problems}
This problem, denoted \rcov, is a {\em two-stage optimization} problem, where elements are possibly bought in the first
stage (at the given cost) or the second stage (at cost $\lambda$ times higher). In the second stage, some subset
$\omega \sse[n]$ of requirements (also called a \emph{scenario}) materializes, and the elements bought in both stages
must collectively satisfy each requirement in $\omega$. Formally, the input to problem \rcov consists of (a) the
covering problem $\cov = \langle E,c, \{\mathcal{R}_i\}_{i=1}^n\rangle$ as above, (b) an uncertainty set $\Omega\sse
2^{[n]}$ of scenarios (possibly implicitly given), and (c)~an inflation parameter $\lambda\ge 1$. A feasible solution
to \rcov is a set of {\em
  first stage elements} $E_0\sse E$ (bought without knowledge of the
scenario), along with an {\em augmentation algorithm} that given any $\omega\in \Omega$ outputs $E_\omega \sse E$ such
that $E_0\cup E_\omega$ satisfies all requirements in $\omega$.  The objective function is  to minimize:
$$c(E_0) + \lambda \cdot \max_{\omega\in\Omega} c(E_\omega).$$
Given such a solution, $c(E_0)$ is called the first-stage cost and
$\max_{\omega\in\Omega} c(E_\omega)$ is the second-stage cost.

Note that by setting $\lambda=1$ in any robust covering problem, \emph{the optimal value of the robust problem equals
that of its corresponding max-min problem}.

As in~\cite{GNR-k-rob}, our algorithms for robust covering problems are based on the following type of guarantee.
In~\cite{GNR-k-rob} these were stated for $k$-robust uncertainty sets, but they immediately extend to arbitrary
uncertainty sets.
\begin{definition}\label{defn:algo}
  An algorithm is \emph{$(\alpha_1,\alpha_2,\beta)$-discriminating} iff
  given as input any instance of $\rcov$ and a threshold $T$, the
  algorithm outputs
  \begin{inparaenum}[(i)]
  \item a set $\fst\sse E$, and
  \item an algorithm $\snd: \Omega\rightarrow
    2^E$,
  \end{inparaenum}
  such that:
  \begin{OneLiners}
  \item[A.] For every scenario $D \in \Omega$,
    \begin{OneLiners}
    \item[(i)] the elements in $\fst~\cup ~\snd(D)$ satisfy all
      requirements in $D$, and
    \item[(ii)] the resulting augmentation cost
      $c\left(\snd(D)\right)\le \beta\cdot T$.
    \end{OneLiners}
  \item[B.] Let $\Phistar$ and $\Tstar$ (respectively) denote the
    first-stage and second-stage cost of an optimal solution to the
    $\rcov$ instance. If the threshold $T\ge \Tstar$ then the first stage
    cost $c(\fst)\le \alpha_1\cdot \Phistar + \alpha_2\cdot \Tstar$.
  \end{OneLiners}
\end{definition}

\begin{lemma}[\cite{GNR-k-rob}]\label{lem:apx}
  If there is an $(\alpha_1,\alpha_2,\beta)$-discriminating algorithm
  for a robust covering problem $\rcov$, then for every $\epsilon > 0$
  there is a $\left((1+\epsilon)\cdot\max\left\{\alpha_1,
    \beta+\frac{\alpha_2}\lambda\right\} \right)$-approximation algorithm for $\rcov$.
\end{lemma}

\subsection{Desirable Properties of the Covering Problem} We now formalize certain properties of the covering problem
$\cov =\langle E,c,\{\rs_i\}_{i=1}^n\rangle$ that are useful in obtaining our results.
Given a partial solution $S\sse E$ and a set $X \sse [n]$ of requirements, any set $E_X \sse E$ such that $S \cup E_X
\in \mathcal{R}_i ~\forall i\in X$ is called an \emph{augmentation} of $S$ for requirements $X$. Given $X, S$, define
the min-cost augmentation of $S$ for requirements $X$ as:
$$\optaug(X\mid S):= \min \{c(E_X)\mid E_X \sse E \text{ and } S\cup E_X\in \mathcal{R}_i,~\forall i\in X\}.$$

Also define $\opt(X) := \min \{c(E_X)\mid E_X \sse E \text{ and } E_X\in \mathcal{R}_i~\forall i\in
X\}=\optaug(X\mid\emptyset)$, for any $X\sse [n]$.

An easy consequence of the fact that costs are non-negative is the following:
\begin{property}[Monotonicity]\label{ass:monotone}
For any requirements $X\sse Y\sse [n]$ and any solution $S\sse E$, $\optaug(X|S)\le \optaug(Y|S)$. Similarly, for any
$X\sse [n]$ and solutions $T\sse S\sse E$, $\optaug(X\mid S) \le \optaug(X\mid T)$.
\end{property}

From the definition of coverage of requirements, we obtain:
\begin{property}[Subadditivity]\label{ass:subadd}
  For any two subsets of requirements $X,Y\sse [n]$ and any partial
  solution $S\sse E$, we have $\optaug(X\mid S) + \optaug(Y\mid S)\ge
  \optaug(X\cup Y\mid S)$.
\end{property}
To see this property: if $\f_X\sse E$ and $\f_Y\sse E$ are solutions corresponding to $\optaug(X\mid S)$ and
$\optaug(Y\mid S)$ respectively, then  $\f_X\cup\f_Y\cup S$ covers requirements $X\cup Y$; so $\optaug(X\cup Y\mid S)
\le c(\f_X\cup \f_Y)\le c(\f_X)+c(\f_Y) = \optaug(X\mid S) + \optaug(Y\mid S)$.

We assume two additional properties of the covering problem:
\begin{property}[Offline Algorithm]\label{ass:apx}
  There is an $\offline$-approximation (offline) algorithm for the
  covering problem $\optaug(X\mid S)$, for any $S\sse E$ and $X\sse [n]$.
\end{property}
\begin{property}[Online Algorithm]\label{ass:online}
  There is a polynomial-time deterministic $\online$-competitive algorithm for the online
  version of $\cov = \langle E,c, \{\mathcal{R}_i\}_{i=1}^n\rangle$.
\end{property}


\subsection{Models of Downward-Closed Families}
All covering functions we deal with are monotone non-decreasing. So we
may assume WLOG that the collection $\Omega$ in both \mmp and \rcov is
\emph{downwards-closed}, i.e. $A\sse B$ and $B\in\Omega$ $\implies$
$A\in\Omega$. In this paper we consider the following well-studied
classes:

\begin{definition}[$p$-system]\label{defn:p-system} A downward-closed family $\Omega\sse 2^{[n]}$ is called a $p$-system iff:
$$ \frac{\max_{I\in \overline{\Omega}, I\sse A}~|I|}{\min_{J\in \overline{\Omega}, J\sse A}~|J|} \le p,\quad
\mbox{ for each }A\sse[n],$$ where $\overline{\Omega}\sse \Omega$ denotes the collection of {\em maximal subsets} in
$\Omega$. Sets in $\Omega$ are called {\em independent sets}. We assume access to a membership-oracle, that given any
subset $I\sse[n]$ returns whether or not $I\in \Omega$.
\end{definition}

\begin{definition}[$q$-knapsack]\label{defn:q-knapsack}
Given $q$ non-negative vectors $w^1,\ldots,w^q: [n]\rightarrow \mathbb{R}_+$ and capacities $b_1,\ldots,b_q\in
\mathbb{R}_+$, the $q$-knapsack constrained family is:
$$\Omega = \left\{ A\sse [n] : \sum_{e\in A} w^j(e)\le b_j, \mbox{ for all }j\in[q]\right\}.$$
\end{definition}

These constraints model a rich class of downward-closed families. Some interesting special cases of $p$-systems are
$p$-matroid intersection~\cite{Schr-book} and $p$-set packing~\cite{HS89,B00}; see the appendix in~\cite{CCPV07} for
more discussion on $p$-systems. Jenkyns~\cite{J76} showed that the natural greedy algorithm is a $p$-approximation for
maximizing linear functions over $p$-systems, which is the best known result. Maximizing a linear function over
$q$-knapsack constraints is the well-studied class of packing integer programs (PIPs), eg.~\cite{S99}. Again, the
greedy algorithm is known to achieve an $O(q)$-approximation ratio. When the number of constraints $q$ is constant,
there is a PTAS~\cite{CK04-multidim}.


\section{Algorithms for Max-Min Optimization}
\label{sec:max-min}

In this section we give approximation algorithms for constrained max-min optimization, i.e. Problem~(\maxf) where $f$
is given by some underlying covering problem and $\I$ is given by some $p$-system and $q$-knapsack. We first consider
the case when $\I$ is a  $p$-system. Then we show that any knapsack constraint can be reduced to a $1$-system
(specifically a partition matroid) in a black-box fashion; this enables us to obtain an algorithm for $\I$ being the
intersection of a  $p$-system and $q$-knapsack. The results of this section assume Properties~\ref{ass:subadd}
and~\ref{ass:online}.

\subsection{Algorithm for $p$-System Constraints}
The algorithm given below is a greedy algorithm, however it is relative to the objective of the online algorithm
$\a_{on}$ from Property~\ref{ass:online} rather than the (approximate) function value itself.
\begin{algorithm}
  \caption{Algorithm for \mmp under $p$-system}
  \begin{algorithmic}[1]
\STATE \textbf{input:} the covering instance \cov that defines $f$ and $p$-system $\Omega$.

\STATE \textbf{let} current scenario $A_0 \gets \emptyset$, counter $i\leftarrow 0$, input sequence $\sigma\gets
\langle \rangle$.

\WHILE{($\exists e \in [n]\setminus A_i$ such that $A_i \cup \{e\} \in \Omega$)}

    \STATE $a_{i+1}\leftarrow \arg\max\left\{ c(\a_{on}(\sigma \circ e)) -
    c(\a_{on}(\sigma)) \,\, : \,\,e\in [n]\setminus A_i \text{ and } A_i \cup
    \{e\}\in \Omega\right\}$.

    \STATE \textbf{let} $\sigma \gets \sigma \circ a_{i+1}$,\,\, $A_{i+1} \gets A_i \cup \{a_{i+1}\}, \,\, i \gets i+1$.

    \ENDWHILE

    \STATE \label{step:mm} \textbf{let} $D \gets A_i$ be the independent set constructed by the above loop.

    \STATE \textbf{output} solution $D$.
\end{algorithmic}
\end{algorithm}

\begin{theorem}\label{thm:maxmin-p-system}
Assuming Properties~\ref{ass:subadd} and~\ref{ass:online} there is a $\left((p+1)\, \online\right)$-approximation
algorithm for \mmp under $p$-systems.
\end{theorem}
\begin{proof}
The proof of this lemma closely follows that in~\cite{CCPV07} for submodular maximization over a $p$-system. We use
slightly more notation that necessary since this proof will be used in the next section as well.

Suppose that the algorithm performed $k$ iterations; let $D=\{a_1,\cdots,a_k\}$ be the ordered set of elements added by
the algorithm. Define $\sigma=\langle\rangle$, $G_0:=\emptyset$, and $G_i:=\a_{on}(\sigma \circ a_1\cdots a_i)$ for
each $i\in[k]$. Note that $G_0\sse G_1\sse\cdots\sse G_k$. It suffices to show that:
\begin{equation}\label{eq:max-min-psystem} \optaug(B\mid G_0)\le (p+1)\cdot c(G_k\setminus G_0) \qquad  \mbox{for every
 } B\in\Omega.\end{equation} This would imply $\opt(B)\le (p+1)\cdot c(G_k)\le (p+1)\,\online \cdot \opt(D)$ for every
$B\in\Omega$, and hence that $D$ is the desired approximation.

We use the following claim proved in~\cite{CCPV07}, Appendix~B (this claim relies on the properties of a $p$-system).

\begin{claim}[\cite{CCPV07}]\label{cl:ccpv}
For any $B\in \Omega$, there is a partition $\{B_i\}_{i=1}^k$ of $B$ such that for all $i\in[k]$,
\begin{enumerate}
 \item $|B_i|\le p$, and
 \item For every $e\in B_i$, we have $\{a_1,\cdots,a_{i-1}\}\bigcup \{e\} \in \Omega$.
\end{enumerate}
\end{claim}

For any sequence $\pi$ of requirements and any $e\in [n]$ define $\aug(e;\pi):=c(\a_{on}(\pi\circ e))-
c(\a_{on}(\pi))$. Note that this function depends on the particular online algorithm. From the second condition in
Claim~\ref{cl:ccpv}, it follows that each element of $B_i$ was a feasible augmentation to $\{a_1,\ldots,a_{i-1}\}$ in
the $i^{th}$ iteration of the {\bf while} loop. By the greedy choice,
\begin{eqnarray}
c(G_i)-c(G_{i-1}) ~=~    \aug(a_i; \sigma \circ a_1\cdots a_{i-1}) & \ge & \max_{e\in B_i}~ \aug(e; \sigma \circ a_1\cdots a_{i-1})\notag\\
    & \ge &\frac{1}{|B_i|}\sum_{e\in B_i} \aug(e; \sigma \circ a_1\cdots a_{i-1})\notag\\
    &\ge & \frac{1}{|B_i|} \sum_{e\in B_i} \optaug(\{e\}\mid G_{i-1}) \label{eq:psys1}\\
    &\ge & \frac{1}{|B_i|} \cdot \optaug(B_i\mid G_{i-1})\label{eq:psys2}\\
    &\ge & \frac{1}{p} \cdot \optaug(B_i\mid G_{i-1}).\label{eq:psys3}
\end{eqnarray}
Above equation~\eqref{eq:psys1} is by the definition of $G_{i-1}= \a_{on}(\sigma \circ a_1\cdots a_{i-1})$,
equation~\eqref{eq:psys2} uses the subadditivity Property~\ref{ass:subadd}, and~\eqref{eq:psys3} is by the first
condition in Claim~\ref{cl:ccpv}.

Summing over all iterations $i\in [k]$, we obtain:
$$c(G_k)- c(G_0)= \sum_{i=1}^k \aug(a_i; \sigma \circ a_1\cdots a_{i-1}) \ge \frac{1}{p} \sum_{i=1}^k
  \optaug(B_i\mid G_{i-1})\ge \frac{1}{p} \sum_{i=1}^k \optaug(B_i\mid G_k)$$ where the last inequality follows from monotonicity since
  $G_{i-1}\sse G_k$ for all $i\in [k]$.

Using subadditivity~\lref[Property]{ass:subadd}, we get $c(G_k)- c(G_0)
  \ge \frac1p \cdot \optaug(\cup_{i=1}^k B_i\mid G_k) = \frac1p \cdot \optaug(B\mid G_k)$.

Let $J:= \arg\min \{c(J')\mid J'\sse E, \mbox{ and }G_k\cup J'\sse
  \mathcal{R}_e,~\forall e\in B\}$. i.e.  $\optaug(B\mid G_k)=c(J)$. Observe that
$J\cup (G_k\setminus G_0)$ is a feasible  augmentation to $G_0$ that covers requirements $B$. Thus,
$$\optaug(B\mid G_0) \le c(J) + c(G_k\setminus G_0) = \optaug(B\mid G_k) + c(G_k\setminus G_0) \le (p+1)\cdot c(G_k\setminus
G_0).$$ This completes the proof.
\end{proof}

\subsection{Reducing knapsack constraints to partition matroids}
\label{sec:knapsack-mat}

In this subsection we show that every knapsack constraint can be reduced to a suitable collection of partition
matroids. This property is then used to complete the algorithm for \mmp when $\Omega$ is given by a $p$-system {\em
and} a $q$-knapsack.
Observe that even a single knapsack constraint need not correspond exactly to a small $p$-system: eg. the knapsack with
weights $w_1=1$ and $w_2=w_3=\cdots=w_n=\frac1n$, and capacity one is only an $(n-1)$-system (since both $\{1\}$ and
$\{2,3,\cdots,n\}$ are maximal independent sets).
However we show that any knapsack constraint can be {\em approximately} reduced to a partition matroid (which is a
$1$-system).
The main idea in this reduction is an enumeration method from Chekuri and Khanna~\cite{CK04-multidim}.

\begin{lemma}\label{lem:knap-to-mat}
  Given any knapsack constraint $\sum_{i=1}^n w_i\cdot x_i\le B$ and
  fixed $0<\epsilon\le 1$, there is a polynomial-time computable
  collection $\p_1,\cdots,\p_T$ of $T=n^{O(1/\epsilon^2)}$ partition matroids such that:
  \begin{enumerate}
  \item For every $X\in \cup_{t=1}^T \p_t$, we have $\sum_{i\in X} w_i
    \le (1+\epsilon)\cdot B$.
  \item $\{X\sse[n] \mid \sum_{i\in X} w_i \le B\}\sse \cup_{t=1}^T
    \p_t$.
  \end{enumerate}
\end{lemma}
\begin{proof}
Let $\delta=\epsilon/6$ and $\beta=\frac{\delta B}n$. WLOG we assume that $\max_{i=1}^n w_i\le B$. Partition the
groundset $[n]$ into $G:=\lceil \frac{\log (n/\delta)}{\log (1+\delta)}\rceil$ groups as follows.
$$
S_k := \left\{
\begin{array}{ll}
\left\{i\in[n] : w_i\le \beta \right\} & \mbox{ if }k=0\\
\left\{i\in[n] : \beta\cdot (1+\delta)^{k-1}<w_i\le \beta \cdot (1+\delta)^{k} \right\} & \mbox{ if } 1\le k\le G\\
\end{array}
\right.
$$
Let $T$ denote the number of non-negative integer partitions of $\lceil G/\delta\rceil$ into $G$ parts. Note that,
$$T:={\lceil G/\delta\rceil+G-1 \choose G-1}\le \exp(\lceil G/\delta\rceil+G-1) \le n^{O(1/\delta^2)}.$$

We will define a collection of $T$ {\em partition matroids} on $[n]$, each over the partition $\{S_0,S_1,\ldots,
S_G\}$. For any integer partition $\tau=\{U_k\}_{k=1}^G$ of $\lceil G/\delta\rceil$ (i.e. $U_k\ge 0$ are integers and
$\sum_k U_k= \lceil G/\delta\rceil$), define a partition matroid $\p_\tau$ that has bounds $N_k(\tau)$ on each part
$S_k$, where
$$
N_k(\tau) := \left\{
\begin{array}{ll}
\infty & \mbox{ if }k=0\\
\lfloor \frac{n\cdot (U_k+1)}{G\cdot (1+\delta)^{k-1}} \rfloor& \mbox{ if } 1\le k\le G\\
\end{array}
\right.
$$

Clearly this collection can be constructed in polynomial time for fixed $\epsilon$. We now show that this collection of
partition matroids satisfies the two properties in the lemma.

{\bf (1)} Consider any $X\sse[n]$ that is feasible for some partition matroid, say $\p_\tau$. The total weight of
elements $X\cap S_0$ is at most $n\cdot \beta\le \delta\cdot B$. For any group $1\le k\le G$, the weight of elements
$X\cap S_k$ is at most:
$$ |X\cap S_k|\cdot  \beta \, (1+\delta)^{k} \le N_k(\tau)\cdot \beta \, (1+\delta)^{k}\le \delta(1+\delta)(U_k+1)\cdot
\frac{B}G$$ Hence the total weight of all elements in $X$ is at most:
\begin{eqnarray*}
\delta B + \delta(1+\delta)\frac{B}G \cdot\left( \sum_{k=1}^G U_k+G\right) & \le & \delta B + \delta(1+\delta)\frac{B}G \cdot\left( \frac{G}\delta +1+G\right) \\
&\le &\delta B + \delta(1+\delta)\frac{B}G \cdot\left( \frac{G}\delta + 2G\right) \\
&\le & \delta B + (1+\delta)\cdot\left( B + 2\delta \,B\right) \\
&\le & B+6\delta\, B.
\end{eqnarray*}
Above we use $\delta\le 1$. Finally since $\delta=\epsilon/6$, we obtain the first condition.

{\bf (2)} Consider  any $Y\sse[n]$ that satisfies the knapsack constraint, i.e. $\sum_{i\in Y} w_i\le B$. We will show
that $Y$ is feasible in $\p_\tau$, for some integer partition $\tau$ of $\lceil G/\delta\rceil$ as above. For each
$1\le k\le G$ let $Q_k$ denote the weight of elements in $Y\cap S_k$, and $U_k$ be the unique integer that satisfies
$U_k \cdot \frac{\delta B}G\le Q_k< (U_k+1) \cdot \frac{\delta B}G$. Define $\tau$ to be the integer partition
$\{U_k\}_{k=1}^G$. We have $\sum_k U_k\le G/\delta$, which follows from the fact $B\ge \sum_k Q_k \ge \frac{\delta B}G
\cdot \sum_k U_k$. By increasing $U_k$s arbitrarily so that they total to $\lceil G/\delta\rceil$, we obtain  a
feasible integer partition $\tau$.
We now claim that $Y$ is feasible for $\p_\tau$. Since each element of $S_k$ has weight at least $\beta\cdot
(1+\delta)^{k-1}$, we have
$$|Y\cap S_k|\le \frac{Q_k}{\beta\,(1+\delta)^{k-1}}\le \frac{(U_k+1)\cdot \delta B/G}{(1+\delta)^{k-1}\cdot \delta
B/n} = \frac{n\cdot (U_k+1)}{G\cdot (1+\delta)^{k-1}}.$$ Since $|Y\cap S_k|$ is integral, we obtain $|Y\cap S_k|\le
\lfloor \frac{n\cdot (U_k+1)}{G\cdot (1+\delta)^{k-1}} \rfloor\le N_k(\tau)$. Thus we obtain the second condition.
\end{proof}

\subsection{Algorithm for $p$-System and $q$-Knapsack Constraints}\label{subsec:mm-knap-psystem}
Here we consider \mmp when $\Omega$ is the intersection of $p$-system \ms and a $q$-knapsack (as in
Definition~\ref{defn:q-knapsack}). The idea is to reduce the $q$-knapsack to a single knapsack (losing factor $\approx
q$), then use Lemma~\ref{lem:knap-to-mat} to reduce the knapsack to a 1-system, and finally apply
Theorem~\ref{thm:maxmin-p-system} on the resulting $p+1$ system. Details appear below.

By scaling weights in the knapsack constraints, we may assume WLOG that each knapsack has capacity exactly one; let
$w^1,\cdots,w^q$ denote the weights in the $q$ knapsack constraints. We also assume WLOG that each singleton element
satisfies the $q$-knapsack; otherwise such elements can be dropped from the groundset.

\begin{algorithm}
\caption{Algorithm for \mmp under $p$-system and $q$-knapsack}
  \begin{algorithmic}[1]
  \STATE Approximate the $q$-knapsack by a single knapsack with weights $\sum_{j=1}^q w^j$ and capacity $q$; applying
  Lemma~\ref{lem:knap-to-mat} with $\epsilon=\frac12$ on this knapsack, let $\{\p_j\}_{j=1}^L$ denote the resulting partition matroids (note $L=n^{O(1)}$).

 \STATE For each $j\in [L]$, define $\Sigma_j := \ms\bigcap \p_j$; note that each $\Sigma_j$ is a $(p+1)$-system.

 \STATE Run the algorithm from Theorem~\ref{thm:maxmin-p-system} under each $p+1$ system $\{\Sigma_j\}_{j=1}^L$ to obtain
solutions $\{E_j\in \Sigma_j\}_{j=1}^L$.

 \STATE Let $j^*\gets \arg\max_{j=1}^L \, c\left(\a_{on}(E_j)\right)$.

 \STATE Partition $E_{j^*}$ into $\{\omega_i\}_{i=1}^{3q+1}$ such that each $\omega_i\in \Omega$, as per Claim~\ref{cl:knap-mat-2}.
 \STATE Output $\omega_{i^*}$ where $i^*\gets \arg\max_{i=1}^{3q+1} \, c\left(\a_{off}(\omega_i)\right)$. Here we use
 the offline algorithm from Property~\ref{ass:apx}.
\end{algorithmic}
\end{algorithm}

We now establish the approximation ratio of this algorithm.

\begin{claim}\label{cl:knap-mat-1}
$\Omega\sse \cup_{j=1}^L \Sigma_j$.
\end{claim}
\begin{proof}
For any $\omega\in \Omega$, we have $\sum_{e\in \omega} w^i(e) \le 1$ for all $i\in [q]$. Hence  $\sum_{e\in \omega}
\sum_{i=1}^q w^i(e) \le q$, i.e. it satisfies the combined knapsack constraint. Now by Lemma~\ref{lem:knap-to-mat}~(2),
we obtain $\omega\in \bigcup_{j=1}^L \p_j$. Finally, since $\omega\in \Omega\sse \ms$, we have  $\omega\in \cup_{j=1}^L
\Sigma_j$.
\end{proof}

\begin{claim}\label{cl:knap-mat-2}
For each $\tau\in \cup_{j=1}^L \Sigma_j$ there exists a collection $\{\omega_i\}_{i=1}^{3q+1}$ such that
$\tau=\bigcup_{\ell=1}^{3q+1} \omega_\ell$, and $\omega_\ell\in \Omega$ for all $\ell\in [3q+1]$. Furthermore, this is
computable in polynomial time.
\end{claim}
\begin{proof}
Consider any $\tau\in \Sigma:=\cup_{j=1}^L \Sigma_j$. Note that $\tau\in\ms$, so any subset of $\tau$ is also in $\ms$
(which is downwards-closed). We will show that there is a partition of $\tau$ into $\{\omega_\ell\}_{\ell=1}^{3q}$ such
that each $\omega_\ell$ satisfies the $q$-knapsack. This suffices to  prove the claim. Since $\tau\in \bigcup_{j=1}^L
\p_j$, by Lemma~\ref{lem:knap-to-mat}~(1) it follows that $\sum_{e\in \tau} \sum_{i=1}^q w^i(e) \le \frac32 q$.
Starting with the trivial partition of $\tau$ into singleton elements, greedily merge parts as long as each part
satisfies the $q$-knapsack, until no further merge is possible. (Note that the trivial partition is indeed feasible
since each element satisfies the $q$-knapsack.) Let $\{\omega_\ell\}_{\ell=1}^{r}$  denote the parts in the final
partition; we will show $r\le 3q+1$ which would prove the claim. Consider forming $\lfloor r/2\rfloor$ pairs from
$\{\omega_\ell\}_{\ell=1}^{r}$ arbitrarily. Observe that for any pair $\{\omega, \omega'\}$, it must be that
$\omega\cup \omega'$ violates {\em some} knapsack; so $\sum_{e\in \omega\cup\omega'} \sum_{i=1}^q w^i(e)>1$. Thus
$\sum_{e\in \tau} \sum_{i=1}^q w^i(e)> \lfloor r/2\rfloor$. On the other hand, $\sum_{e\in \tau} \sum_{i=1}^q w^i(e)\le
\frac32 q$, which implies $r< 3q+2$.
\end{proof}

\begin{theorem}\label{thm:max-min-psystem-knapsack}
Assuming Properties~\ref{ass:subadd},~\ref{ass:apx} and~\ref{ass:online}, there is an $O((p+1)\,
(q+1)\,\offline\,\online)$-approximation algorithm for \mmp under a $p$-system and $q$-knapsack constraint.
\end{theorem}
\begin{proof}
Let $\opt_j$ denote the optimal value of \mmp under $p+1$ system $\Sigma_j$, for each $j\in [L]$. By
Claim~\ref{cl:knap-mat-1} we have $\max_{j=1}^L \opt_j\ge \opt$, the optimal value of \mmp under $\Omega$. Observe that
Theorem~\ref{thm:maxmin-p-system} actually implies $c(\a_{on}(E_j))\ge \frac1{p+2}\cdot \opt_j$ for each $j\in[q]$.
Thus $c(\a_{on}(E_{j^*}))\ge \frac1{p+2}\cdot \opt$; hence $\opt(E_{j^*}) \ge \frac1{\online\,(p+2)}\cdot \opt$. Now
consider the partition $\{\omega_i\}_{i=1}^{3q+1}$ of $E_{j^*}$ from Claim~\ref{cl:knap-mat-2}. By the subadditivity
property, $\sum_{i=1}^{3q+1} \opt(\omega_i) \ge \opt(E_{j^*})$; i.e. there is some $i'\in[3q+1]$ with
$\opt(\omega_{i'})\ge \frac1{\online\, (p+2)(3q+1)}\cdot \opt$. Thus the $i^*$ found using the offline algorithm
(Property~\ref{ass:apx}) satisfies $\opt(\omega_{i^*})\ge \frac1{\online\, \offline\,(p+2)(3q+1)}\cdot \opt$.
\end{proof}

{\bf Remark:} We can obtain a better approximation guarantee of $O((p+1)\, (q+1)\,\online)$ in
Theorem~\ref{thm:max-min-psystem-knapsack} using randomization. This algorithm is same as Algorithm~2, except for the
last step, where we output $\omega_{\ell}$ for $\ell\in[3q+1]$ chosen {\em uniformly at random}. From the above proof
of Theorem~\ref{thm:max-min-psystem-knapsack}, it follows that:
$$E[\opt(\omega_{\ell})] = \frac1{3q+1} \sum_{i=1}^{3q+1} \opt(\omega_i) \ge \frac{\opt(E_{j^*})}{3q+1}\ge \frac1{\online\,
(p+2)(3q+1)}\cdot \opt.$$

\section{General Framework for Robust Covering Problems} \label{sec:gen-sets}
\newcommand{\mapx}{\alpha_{\sf mm}}

In this section we present an abstract framework for robust covering problems under {\em any uncertainty set} $\Omega$,
as long as we are given access to offline, online and max-min algorithms for the base covering problem. Formally, this
requires Properties~\ref{ass:apx},~\ref{ass:online} and the following additional property (recall the notation from
Section~\ref{sec:prelim}).
\begin{property}[Max-Min Algorithm]\label{ass:maxmin}
  There is an $\mapx$-approximation algorithm for the max-min problem:
  given input $S\sse E$, $\mm(S):= \max_{X\in\Omega} \min \{c(A)\mid S\cup
  A\in R_i,~\forall i\in X\}$.
\end{property}

\begin{theorem}\label{th:gen-p-sets}
Under \lref[Properties]{ass:subadd},~\ref{ass:apx},~\ref{ass:online} and \ref{ass:maxmin}, there is an
$O(\offline\cdot\online\cdot\mapx)$-approximation algorithm for the robust covering problem $\rcov = \langle E,c,
\{R_i\}_{i=1}^n, \Omega,\lambda \rangle$.
\end{theorem}
\begin{proof}
The algorithm proceeds as follows.
\begin{algorithm}
  \caption{Algorithm Robust-with-General-Uncertainty-Sets}
  \begin{algorithmic}[1]
\STATE \textbf{input:} the \rcov instance and threshold $T$.

    \STATE \textbf{let} counter $t\leftarrow 0$, initial online
    algorithm's input $\sigma = \langle\rangle$, initial online solution
    $F_0\leftarrow \emptyset$.

    \REPEAT

      \STATE {\bf set} $t\leftarrow t+1$.

    \STATE \label{step:gen:1} \textbf{let} $E_t\sse [n]$ be the scenario returned by
    the algorithm of \lref[Property]{ass:maxmin} on $\mm(F_{t-1})$.

    \STATE \label{step:gen:2} \textbf{let} $\sigma \gets \sigma \circ E_t$, and $F_t \gets
    \a_{on}(\sigma)$ be the current online solution.

    \UNTIL{$c(F_t) - c(F_{t-1}) \le 2\online \cdot T$} \label{step:gen:3}

\STATE  \textbf{set} $\tau \leftarrow t-1$.

    \STATE \textbf{output} first-stage solution $\fst:=F_{\tau}$.
    \STATE \textbf{output} second-stage solution $\snd$ where for any $\omega\sse[n]$, $\snd(\omega)$ is the solution of
  the offline algorithm (\lref[Property]{ass:apx}) for the problem $\optaug(\omega\mid \fst)$.
  \end{algorithmic}
\end{algorithm}

As always, let $\Phistar \sse E$ denote the optimal first stage solution (and its cost), and $\Tstar$ the optimal
  second-stage cost; so the optimal value is $\Phistar+\lambda\cdot \Tstar$. We prove the
  performance guarantee using the following claims.
\begin{Myquote}
\begin{claim}[General 2nd stage]\label{cl:p1}
For any $T\ge 0$ and $X\in\Omega$, elements $\fst\bigcup \snd(X)$ satisfy all the requirements in $X$, and
  $c(\snd(X))\le 2\offline\cdot \mapx\cdot \online\cdot T$.
\end{claim}
\begin{proof}
It is clear that $\fst\bigcup \snd(X)$ satisfy all requirements in $X$. By the choice of set $E_{\tau+1}$ in
\lref[line]{step:gen:1} of the last iteration, for any $X\in\Omega$ we have:
$$\optaug(X\mid F_\tau)\le \mapx\cdot \optaug(E_{\tau+1}\mid F_\tau)\le \mapx\cdot \left( c(F_{\tau+1}) -
c(F_{\tau})\right)\le 2\mapx\cdot \online\cdot T$$ The first inequality is by \lref[Property]{ass:maxmin}, the second
inequality uses the fact that $F_{\tau+1}\supseteq F_\tau$ (since we use an online algorithm to augment in
\lref[line]{step:gen:2}),\footnote{This is the technical reason we need an online algorithm. If instead we had used an
offline algorithm to compute $F_t$ in step~\ref{step:gen:2} then $F_t\not\supseteq F_{t-1}$ and we could not upper
bound the augmentation cost $\optaug(E_t\mid F_{t-1})$ by $c(F_t)-c(F_{t-1})$.} and the last inequality follows from
the termination condition in \lref[line]{step:gen:3}. Finally, since $\snd(X)$ is an $\offline$-approximation to
$\optaug(X\mid F_\tau)$, we obtain the claim.
\end{proof}

\begin{claim}\label{cl:p2}
$\opt(\cup_{t\le \tau} E_t)\le \tau\cdot \Tstar+ \Phistar$.
\end{claim}
\begin{proof}
Since each $E_t\in \Omega$ (these are solutions to $\mm$), the bound on the second-stage optimal cost gives
$\optaug(E_t\mid \Phistar)\le \Tstar$ for all $t\le \tau$. By subadditivity (\lref[Property]{ass:subadd}) we have
$\optaug(\cup_{t\le \tau} E_t\mid \Phistar)\le \tau\cdot \Tstar$, which immediately implies the claim.
\end{proof}

\begin{claim}\label{cl:p3}
$\opt(\cup_{t\le \tau} E_t)\ge \frac1\online \cdot c(F_\tau)$.
\end{claim}
\begin{proof}
Directly from the competitiveness of the online algorithm in \lref[Property]{ass:online}.
\end{proof}

\begin{claim}[General 1st stage]\label{cl:p4}
If $T\ge \Tstar$ then $c(\fst)=c(F_\tau)\le 2\,\online\cdot \Phistar$.
\end{claim}
\begin{proof}
We have $c(F_\tau)= \sum_{t=1}^\tau \left[ c(F_t)- c(F_{t-1})\right] > 2\online\tau\cdot T\ge 2\online\tau\cdot \Tstar$
by the choice in Step~\eqref{step:gen:3}. Combined with \lref[Claim]{cl:p3}, we have $\opt(\cup_{t\le \tau} E_t)\ge
2\tau\cdot \Tstar$. Now using \lref[Claim]{cl:p2}, we have $\tau\cdot \Tstar\le \Phistar$, and hence $\opt(\cup_{t\le
\tau} E_t)\le 2\cdot \Phistar$. Finally using \lref[Claim]{cl:p3}, we obtain $c(F_\tau)\le 2\online\cdot \Phistar$.
\end{proof}
\end{Myquote}

\lref[Claim]{cl:p1} and \lref[Claim]{cl:p4} imply that the above algorithm is a
$(2\online,~0,~2\mapx\online\offline)$-discriminating algorithm for the robust problem $\rcov=\langle E,c,
\{R_i\}_{i=1}^n, \Omega,\lambda \rangle$. Now using \lref[Lemma]{lem:apx} we obtain the theorem.
\end{proof}

\paragraph{Explicit uncertainty sets} An easy consequence of \lref[Theorem]{th:gen-p-sets} is for the {\em  explicit scenario} model of robust covering
problems~\cite{DGRS05,GGR06}, where $\Omega$ is specified as a list of possible scenarios. In this case, the \mm
problem can be solved using the $\offline$-approximation algorithm from \lref[Property]{ass:apx} which implies an
$O(\offline^2\online)$-approximation for the robust version. In fact, we can do slightly better---observing that in
this  case, the algorithm for second-stage augmentation is the same as the Max-Min algorithm, we obtain an
$O(\offline\cdot \online)$-approximation algorithm for robust covering with explicit scenarios. As an application of
this result, we obtain an $O(\log n)$ approximation for robust Steiner forest with explicit scenarios, which is the
best known result for this problem.

\section{Robust Covering under $p$-System and $q$-Knapsack Uncertainty
  Sets}
\label{sec:combine}

Recall that any uncertainty set $\Omega$ for a robust covering problem can be assumed WLOG to be {\em downward-closed},
i.e. $X\in \Omega$ and $Y\sse X$ implies $Y\in \Omega$. Eg., in the $k$-robust model $\Omega=\{S\sse[n] : |S|\le k\}$.
Hence it is of interest to obtain good approximation algorithms for robust covering when $\Omega$ is specified by means
of general models for downward-closed families. In this section, we consider the two well-studied models of $p$-systems
and $q$-knapsacks (Definitions~\ref{defn:p-system} and~\ref{defn:q-knapsack}).

\ignore{In this section, we consider robust covering problems under uncertainty sets described via $p$-systems and
knapsack constraints. Recall the framework defined in \lref[Section]{sec:notation}: we have a covering problem \cov
with $m$ requirements, and the possible second-stage scenarios are given by some downward-closed $\Omega \sse 2^{[n]}$.
Here we consider uncertainty sets specified by means of the following two well-studied models for downward-closed
families.}

The result of this section says the following: \emph{if we can solve
  both the offline and online versions of a covering problem well, we
  get good algorithms for \rcov under uncertainty sets given by the
  intersection of $p$-systems and $q$-knapsack constraints}. Naturally, the
performance depends on $p$ and $q$; we note  that this is unavoidable due to complexity considerations. Based on
Theorem~\ref{th:gen-p-sets} it suffices to give an approximation algorithm for the max-min problem under $p$-systems
and $q$-knapsack constraints; so Theorem~\ref{thm:max-min-psystem-knapsack} combined with Theorem~\ref{th:gen-p-sets}
implies an $O\left((p+1)(q+1)\,\online^2\,\offline^2\right)$-approximation ratio. However, we can obtain a better
guarantee by considering the algorithm for \rcov directly. Formally we show that:
\begin{theorem}
  \label{th:general}
  Under \lref[Properties]{ass:subadd},~\ref{ass:apx} and
  \ref{ass:online}, the robust covering problem $\rcov\langle E,c,
  \{\mathcal{R}_i\}_{i=1}^m, \Omega,\lambda \rangle$ admits an $O\left(
    (p+1)\cdot(q+1)\cdot \offline\cdot\online\right)$-approximation
  guarantee when $\Omega$ is given by the intersection of a $p$-system and $q$-knapsack constraints.
\end{theorem}
The outline of the proof is same as for Theorem~\ref{thm:max-min-psystem-knapsack}. We first consider the case when the
uncertainty set is a $p$-system (\lref[subsection]{subsec:framework-mat}); then using the reduction in
Lemma~\ref{lem:knap-to-mat} we solve a suitable instance of \rcov under a $(p+1)$-system uncertainty set.

\subsection{$p$-System Uncertainty Sets}
\label{subsec:framework-mat} In this subsection, we consider \rcov when the uncertainty set $\Omega$ is some
$p$-system. The algorithm is a combination of the ones in Theorem~\ref{th:gen-p-sets} and
Theorem~\ref{thm:maxmin-p-system}.  We start with an empty solution, and use the online algorithm to greedily try and
build a scenario of large cost. If we do find a ``violated'' scenario which is unhappy with the current solution, we
augment our current solution to handle this scenario (again using the online algorithm), and continue. The algorithm is
given as Algorithm~4 below.

\begin{algorithm}
  \caption{Algorithm Robust-with-$p$-system-Uncertainty-Sets }
  \begin{algorithmic}[1]
  \STATE \textbf{input:} the \rcov instance and bound $T$.

    \STATE \textbf{let} counter $t\leftarrow 0$, initial online
    algorithm's input $\sigma = \langle\rangle$, initial online solution
    $F_0\leftarrow \emptyset$.

    \REPEAT

  \STATE {\bf set} $t\leftarrow t+1$.

    \STATE \textbf{let} current scenario $A^t_0 \gets \emptyset$,
    counter $i\leftarrow 0$.

    \WHILE{($\exists e \in [n]\setminus A^t_i$ such that $A^t_i \cup
      \{e\} \in \Omega$)}

    \STATE $a_{i+1}\leftarrow \arg\max\{ c(\a_{on}(\sigma \circ e)) -
    c(\a_{on}(\sigma)) \mid e\in [n]\setminus A_i \text{ and } A_i \cup
    \{e\}\in \Omega\}$.

    \STATE \textbf{let} $\sigma \gets \sigma \circ a_{i+1}$,
    $A^t_{i+1} \gets A^t_i \cup \{a_{i+1}\}, i \gets i+1$.

    \ENDWHILE

    \STATE \label{step:mm} \textbf{let} $E_t \gets A^t_i$ be the
    scenario constructed by the above loop.

    \STATE \label{step:online} \textbf{let} $F_t \gets
    \a_{on}(\sigma)$ be the current online solution.

    \UNTIL{$c(F_t) - c(F_{t-1}) \le 2\online \cdot
      T$} \label{step:aug}

\STATE  \textbf{set} $\tau \leftarrow t-1$.

    \STATE \textbf{output} first-stage solution $\fst:=F_{\tau}$.

  \STATE \textbf{output} second-stage solution $\snd$ where for any $\omega\sse[n]$, $\snd(\omega)$ is the solution of
  the offline algorithm (Property~\ref{ass:apx}) for the problem $\optaug(\omega\mid \fst)$.

\end{algorithmic}
\end{algorithm}

We first prove a useful lemma about the behavior of the \textbf{while} loop. \ignore{ though it does not return a
scenario whose cost of augmentation with respect to the current solution is the largest possible, the loop does give a
scenario which is not much worse. This key lemma is proved in the next subsection.}
\begin{lemma}[Max-Min Lemma]
  \label{lem:maxmin}
  For any iteration $t$ of the {\bf repeat} loop, the scenario $E_t \in \Omega$ has the property that for any
  other scenario $B \in \Omega$, $\optaug(B\mid F_{t-1}) \le (p+1)\cdot c(F_t\setminus F_{t-1})$.
\end{lemma}
\begin{proof}
The proof is almost identical to that of Theorem~\ref{thm:maxmin-p-system}.

Consider any iteration $t$ of the {\bf repeat} loop in  Algorithm~4  that starts with a sequence $\sigma$  of elements
(that have been fed to the online algorithm
  $\a_{on}$). Let $A=\{a_1,\cdots,a_k\}$ be the ordered set of elements
  added by the algorithm in this iteration. Define
  $G_0:=\a_{on}(\sigma)$, and $G_i:=\a_{on}(\sigma \circ a_1\cdots a_i)$
  for each $i\in[k]$. Note that $F_{t-1}=G_0$ and $F_t=G_k$, and
  $G_0\sse G_1\sse\cdots\sse G_k$. It suffices to show that
  $\optaug(B\mid G_0)\le (p+1)\cdot c(G_k\setminus G_0)$ for every
  $B\in\Omega$. But this is precisely Equation~\eqref{eq:max-min-psystem} from the proof of Theorem~\ref{thm:maxmin-p-system}.
\end{proof}

\begin{corollary}[Second Stage]\label{cor:mk-2nd}
  For any $T\ge 0$ and $B\in\Omega$, elements $\fst\bigcup \snd(B)$ satisfy all the requirements in $B$, and
  $c(\snd(B))\le 2\offline\cdot \online\cdot (p+1)\cdot T$.
\end{corollary}
\begin{proof}
Observe that $\fst=F_\tau=\a_{on}(\sigma)$, so the first part of the corollary follows from the definition of \snd.  By
\lref[Lemma]{lem:maxmin} and the termination condition on
  \lref[line]{step:aug}, we have $\optaug(B\mid F_\tau)\le (p+2)\cdot  (c(F_{\tau+1}) - c(F_{\tau})) \leq 2(p+2)\online\, T$. Now
  \lref[Property]{ass:apx} guarantees that the solution $\snd(B)$ found by this approximation algorithm has cost at most
$2\offline\cdot \online\cdot(p+2)\, T$.
  \end{proof}

It just remains to bound the cost of the first-stage solution $F_{\tau}$. Below $\Phistar$ denotes the optimal
first-stage solution (and its cost); and $\Tstar$ is the optimal second-stage cost.
\begin{lemma}[First Stage]\label{lem:mk-1st}
If $T\ge \Tstar$ then $c(\fst)=c(F_\tau) \leq  2\online\cdot\Phistar$.
\end{lemma}
\begin{proof}
For any set $X\sse[n]$ of requirements let $\opt(X)$ denote the minimum cost to satisfy $X$. Firstly, observe that
$\opt(\cup_{t\le \tau} E_t)\le \tau\cdot \Tstar+\Phistar$.
  This follows from the fact that each of the $\tau$ scenarios $E_t$ are
  in $\Omega$, so the bound on the second-stage optimal cost gives
  $\optaug(E_t\mid \Phistar)\le \Tstar$ for all $t\le \tau$. By
  subadditivity (\lref[Assumption]{ass:subadd}) we have
  $\optaug(\cup_{t\le \tau} E_t\mid \Phistar)\le \tau\cdot \Tstar$,
  which immediately implies the inequality. Now, we claim that
  \begin{equation}
    \label{mk:eq2}
    \ts \opt(\cup_{t\le \tau} E_t)\ge \frac1\online \cdot c(F_\tau) \geq
    \frac{1}{\online} \cdot 2\online\tau\cdot \Tstar = 2\tau\cdot \Tstar.
  \end{equation}
  The first inequality follows directly from the competitiveness of the
  online algorithm in \lref[Assumption]{ass:online}. For the second
  inequality, we have $c(F_\tau)= \sum_{t=1}^\tau \left[ c(F_t)-
    c(F_{t-1})\right] > 2\online\tau\cdot T\ge 2\online\tau\cdot \Tstar$ by the terminal
  condition in \lref[Step]{step:aug}.  Putting~ the upper and lower bounds on
 $\opt(\cup_{t\le \tau} E_t)$ together, we have $\tau\cdot \Tstar\le \Phistar$,
  and hence $\opt(\cup_{t\le \tau} E_t)\le 2\cdot
 \Phistar$. Using the competitiveness of the online
  algorithm again, we obtain  $c(F_\tau)\le 2\online\cdot \Phistar$.
\end{proof}

From \lref[Corollary]{cor:mk-2nd} and \lref[Lemma]{lem:mk-1st}, it follows that our algorithm is $\left(2\online, 0,
2\offline\,\online\cdot (p+1) \right)$-discriminating (cf. Definition~\ref{defn:algo}) to \rcov. Thus we obtain
Theorem~\ref{th:general} for the case $q=0$.

\ignore{outputs a solution of objective value at most $2\online\cdot \Phistar + 2\offline\online(p+2)\cdot
\lambda\Tstar$, where the second stage augmentation algorithm is just the offline approximation from
\lref[Property]{ass:apx}. This completes the proof of \lref[Theorem]{th:general} for the case of just matroid
constraints (i.e. $p = 0$). In \lref[Subsection]{sec:knapsack-mat}, we show how knapsack constraints can be reduced to
partition matroids, so as to obtain \lref[Theorem]{th:general}.}

\subsection{Algorithm for $p$-Systems {\em and } $q$-Knapsacks }
Here we consider \rcov when the uncertainty set $\Omega$ is the intersection of   $p$-system \ms and a $q$-knapsack.
The algorithm is similar to that in Subsection~\ref{subsec:mm-knap-psystem}. Again, by scaling weights in the knapsack
constraints, we may assume WLOG that each knapsack has capacity exactly one; let $w^1,\cdots,w^q$ denote the weights in
the $q$ knapsack constraints. We also assume WLOG that each singleton element satisfies the $q$-knapsack. The algorithm
for \rcov under $\Omega$ works as follows.

\begin{algorithm}
\caption{Algorithm Robust with $p$-system and $q$-knapsack Uncertainty Set}
  \begin{algorithmic}[1]
  \STATE Consider a modified uncertainty set $\Omega'$ that is given by the intersection of \ms and the {\em single knapsack}
with weight-vector $\sum_{j=1}^q w^j$ and capacity $q$.

  \STATE Applying the algorithm in Lemma~\ref{lem:knap-to-mat} to  this single knapsack with $\epsilon=1$, let
$\{\p_j\}_{j=1}^L$ denote the resulting partition matroids (note $L=n^{O(1)}$).

  \STATE For each $j\in [L]$, define uncertainty-set $\Sigma_j := \ms\bigcap \p_j$; note that each $\Sigma_j$ is a
$(p+1)$-system.

  \STATE Let  $\Sigma\gets \cup_{j=1}^L \Sigma_j$. Solve \rcov under $\Sigma$ using the algorithm of
  Theorem~\ref{th:p-sets-union}.
\end{algorithmic}
\end{algorithm}

Recall Claims~\ref{cl:knap-mat-1} and~\ref{cl:knap-mat-2} which hold here as well.

\begin{lemma}\label{lem:knap-mat-combine}
Any $\alpha$-approximate solution to \rcov under $\Sigma$ is a $(3q+1)\alpha$-approximate solution to \rcov under
uncertainty-set $\Omega$.
\end{lemma}
\begin{proof}
Consider the optimal first-stage solution $\Phistar$ to \rcov under $\Omega$, let \Tstar denote the optimal
second-stage cost and \opt the optimal value. Let $\tau\in \Sigma$ be any scenario, with partition
$\{\omega_i\}_{i=1}^{3q+1}$ given by Claim~\ref{cl:knap-mat-2}. Using the subadditivity Property~\ref{ass:subadd}, we
have $\optaug(\tau|\Phistar)\le \sum_{\ell=1}^{3q+1} \optaug(\omega_\ell|\Phistar) \le (3q+1)\cdot \Tstar$. Thus the
objective value of $\Phistar$ for \rcov under $\Sigma$ is at most $c(\Phistar)+\lambda\cdot (3q+1)\, \Tstar \le
(3q+1)\cdot \opt$.

Claim~\ref{cl:knap-mat-1} implies that for any solution, the objective value of \rcov under $\Omega$ is at most that of
\rcov under $\Sigma$. Thus the lemma follows.
\end{proof}

For solving \rcov under $\Sigma$, note that although $\Sigma$ itself is not any $p'$-system, it is the {\em union  of
polynomially-many} $(p+1)$-systems. We show below that a simple extension of the algorithm in
Subsection~\ref{subsec:framework-mat} also works for unions of $p$-systems; this would solve \rcov under $\Sigma$.
\begin{theorem}\label{th:p-sets-union}
There is an $O((p+1)\, \offline\, \online)$-approximation for \rcov when the uncertainty set is given by the  union of
polynomially-many $p$-systems.
\end{theorem}
\begin{proof}
Let $\Sigma=\cup_{j=1}^L \Sigma_j$ denote the uncertainty set where each $\Sigma_j$ is a $p$-system. The algorithm for
\rcov under $\Sigma$ is just Algorithm~4 where we replace the body of the repeat-loop (ie. lines 4-11) by:
  \begin{algorithmic}[1]
  \STATE {\bf set} $t\leftarrow t+1$.

  \FOR{($j\in[L]$)}

  \STATE \textbf{let} current scenario $A_j \gets \emptyset$,

 \WHILE{($\exists e \in [n]\setminus A_j$ such that $A_j \cup
      \{e\} \in \Sigma_j$)}

    \STATE $e^* \leftarrow \arg\max\{ c(\a_{on}(\sigma \circ A_j\circ e)) -
    c(\a_{on}(\sigma\circ A_j)) \mid e\in [n]\setminus A_j \text{ and } A_j \cup
    \{e\}\in \Omega\}$.

    \STATE  $A_{j} \gets A_j \cup \{e^* \}$.

    \ENDWHILE
 \STATE Let $\Delta_j  \gets c(\a_{on}(\sigma\circ A_j)) - c(\a_{on}(\sigma))$.

 \ENDFOR

    \STATE \textbf{let} $j^* \gets \arg\max\{\Delta_j \mid j\in [L]\}$, and $E_t\gets A_{j^*}$.

    \STATE \textbf{let} $\sigma\gets \sigma\circ E_t$ and  $F_t \gets
    \a_{on}(\sigma)$ be the current online solution.
\end{algorithmic}

Consider any iteration $t$ of the repeat loop. By Lemma~\ref{lem:maxmin} applied to each $p$-system $\Sigma_j$,
\begin{claim}
For each $j\in [L]$, we have $\optaug(B|F_{t-1}) \le (p+1)\cdot \Delta_j$ for every $B\in \Sigma_j$.
\end{claim}
By the choice of scenario $E_t$ and since $\Sigma=\bigcup_{j=1}^L \Sigma_j$, we obtain:
\begin{claim}
For any iteration $t$ of the repeat loop and any $B\in \Sigma$, $\optaug(B|F_{t-1}) \le (p+1)\cdot c(F_t\setminus
F_{t-1})$.
\end{claim}
Based on these claims and proofs identical to Corollary~\ref{cor:mk-2nd} and Lemma~\ref{lem:mk-1st}, we obtain the same
bounds on the first and second stage costs of the final solution $F_\tau$. Thus our algorithm is $\left(2\online, 0,
2\offline\,\online\cdot (p+1) \right)$-discriminating, which by Lemma~\ref{lem:apx} implies the theorem.
\end{proof}

Finally, combining Lemma~\ref{lem:knap-mat-combine} and Theorem~\ref{th:p-sets-union} we obtain
Theorem~\ref{th:general}.

\ignore{Using subadditivity, we can reduce the $p$ knapsack constraints to just one knapsack (with ), at the loss of a
factor $p$ in the objective value of the robust problem. Furthermore, using \lref[Lemma]{lem:knap-to-mat}, one can
reduce (in $m^{O(1)}$ time) this knapsack constraint to a partition matroid; this reduction loses an additional $O(1)$
factor in the objective value. Hence the original uncertainty set with $p$ knapsack and $q$ matroid constraints reduces
to one having $q+1$ matroid constraints, at the loss of an $O(p)$ factor. Finally we can use the algorithm for the
matroid-constrained case from the previous subsection, to obtain an $O((p+1)\cdot (q+1)\cdot\offline\cdot \online)$
approximation for robust covering under matroid and knapsack constraints. This completes the proof of
\lref[Theorem]{th:general}.}

{\bf Remark:} In \lref[Theorem]{th:general}, the dependence on the number of constraints describing the uncertainty set
$\Omega$ is inevitable (under some complexity assumptions). Consider a very special case of the robust covering problem
on ground-set $E$, requirements $E$ (where $i\in E$ is satisfied iff the solution contains $i$), a unit cost function
on $E$, inflation parameter $\lambda=1$. The uncertainty set $\Omega$  is given by the intersection of $p$ different
cardinality constraints coming from some {\em set packing} instance on $E$. In this case, the optimal value of the
robust covering problem is exactly the optimal value of the set packing instance. The hardness result from
H{\aa}stad~\cite{H99} now implies that this robust covering problem is $\Omega(p^{\frac12 -\epsilon})$ hard to
approximate. We note that this hardness applies only to algorithms having running time that is sub-exponential in both
$|E|$ and $p$; this is indeed the case for our algorithm.

\ignore{{\bf Remark 2:} Instead of reducing all the knapsacks to a single partition matroid and losing a factor $O(q)$
as above, we can reduce each knapsack to a partition matroid (using \lref[Lemma]{lem:knap-to-mat}). This only incurs an
$O(1)$ factor loss, however this reduction has running time $m^{O(q)}$. After this reduction, we are left with an
uncertainty set that is a $(q+p)$-system for which the algorithm of \lref[Section]{subsec:framework-mat} applies. Thus
there is an $m^{O(q)}$ time $O\left( (p+q+1)\cdot \offline\cdot \online\right)$-approximation algorithm for \rcov under
$q$ matroid and $p$ knapsack constraints. This reduction also implies a $(p+q+1+\epsilon)$-approximation algorithm for
(monotone) submodular maximization subject to a $p$-system and $q$ knapsacks (when $q$ is a constant)...}

\paragraph{Results for $p$-System and $q$-Knapsack Uncertainty Sets.} We now list some specific results for robust covering
under uncertainty sets described by $p$-systems and knapsack constraints; these follow directly from
\lref[Theorem]{th:general} using known offline and (deterministic) online algorithms for the relevant problems.

\begin{center}
\begin{small}
\begin{tabular}{|c|c|c|c|} \hline
{\bf Problem} & {\bf Offline ratio }& {\bf Online ratio} & {\bf $p$-system, $q$-knapsack Robust}\\
\hline\hline
Set Cover & $O(\log m)$ & $O(\log m\cdot \log n)$~\cite{AAABN03} & $pq\cdot \log^2 m\cdot \log n$ \\
\hline
Steiner Tree/Forest & 2~\cite{AKR95,GW95} & $O(\log n)$~\cite{IM91,BC97} & $pq\cdot \log n$\\
\hline Minimum Cut & 1 & $O(\log^3 n\cdot
  \log\log n)$~\cite{AAABN04,HHR03} & $pq\cdot \log^3 n\cdot \log\log n$\\
\hline Multicut & $\log n$~\cite{GVY96} & $O(\log^3 n\cdot  \log\log n)$~\cite{AAABN04,HHR03} & $pq\cdot \log^4 n\cdot \log\log n$\\
\hline
\end{tabular}
\end{small}
\end{center}


\section{Non-Submodularity of Some Covering Functions}
\label{sec:lbd} In this section we show that some natural covering functions are not even approximately submodular. Let
$f:2^U\rightarrow \mathbb{R}_{\ge 0}$ be any monotone subadditive function. We say that $f$ is $\alpha$-approximately
submodular iff there exists a submodular function $g:2^U\rightarrow \mathbb{R}_{\ge 0}$ with $g(S)\le f(S)\le
\alpha\cdot g(S)$ for all $S\sse U$.

Consider the min-set-cover function, $f_{SC}(S)=$ minimum number of sets required to cover elements $S$.
\begin{proposition}
The min-set-covering function is not $o(n)$-approximately submodular.
\end{proposition}
\begin{proof}
The proof follows from the lower bound on {\em budget-balance} for {\em cross-monotone cost allocations}. Immorlica et
al.~\cite{IMM08} showed that there is no $o(n)$-approximately budget-balanced cross-monotone cost allocation for the
set-cover game. On the other hand it is known (see Chapter 15.4.1 in~\cite{AlgGameThy-book}) that any submodular-cost
game admits a budget-balanced cross-monotone cost allocation. This also implies that any $\alpha$-approximately
submodular cost function (non-negative) admits an $\alpha$-approximate budget-balanced cross-monotone cost allocation.
Thus the min-set-covering function can not be $o(n)$-approximately submodular.
\end{proof}

Similarly, for minimum multicut ($f_{MMC}(S)=$ minimum cost cut separating the pairs in $S$),
\begin{proposition}
The min-multicut function is not $o(n^{1/3})$-approximately submodular.
\end{proposition}
\begin{proof}
This uses the result that the vertex-cover game does not admit $o(n^{1/3})$-approximately budget-balanced
cross-monotone cost allocations~\cite{IMM08}. Since multicut (even on a star graph) contains the vertex-cover problem,
the proposition follows.
\end{proof}

On the other hand, some other covering functions are indeed approximately submodular.
\begin{itemize}
\item The minimum-cut function ($f_{MC}(S)=$ minimum cost cut separating vertices $S$ from the root) is in fact submodular due to
submodularity of cuts in graphs.
\item The min-Steiner-tree ($f_{ST}(S)=$ minimum length tree that connects vertices
$S$ to the root) and min-Steiner-forest ($f_{SF}(S)=$ minimum length forest connecting the pairs in $S$) functions are
$O(\log n)$-approximately submodular. When the underlying metric is a tree, these functions are submodular---in this
case they reduce to weighted coverage functions. Using probabilistic approximation of general metrics by trees, we can
write $g(S)=E_{T\in \mathcal{T}}[f^{T}(S)]$ where $\mathcal{T}$ is the distribution on dominating tree-metrics
(from~\cite{FRT03}) and $f^T$ is the Steiner-tree/Steiner-forest function on tree $T$. Clearly $g$ is submodular. Since
there exists $\mathcal{T}$ that approximates distances in the original metric within factor $O(\log n)$~\cite{FRT03},
it follows that $g$ also $O(\log n)$-approximates $f_{ST}$ (resp. $f_{SF}$).
\end{itemize}

While approximate submodularity of the covering problem \cov  (eg. minimum-cut or Steiner-tree) yields direct
approximation algorithms for \mmp, it is unclear whether they help in solving \rcov (even under cardinality-constrained
uncertainty sets~\cite{GNR-k-rob}). On the other hand, the online-algorithms based approach in this paper solves both
\mmp and \rcov, for uncertainty sets from $p$-systems and $q$-knapsacks.


\bibliography{../robust,../../abbrev,../../my-papers,../../embedding}
\bibliographystyle{plain}

\end{document}